# Quantification and Physics of Cold Plasma Treatment of Organic Liquid Surfaces

Edward Bormashenko [a,b*], Victor Multanen [b], Gilad Chaniel [a,c], Roman Grynyov[a], Evgeny Shulzinger[a], Roman Pogreb[a], Hadas Aharoni[b], Yakir Nagar[d]

[a]*Ariel University, Physics Faculty, 40700, P.O.B. 3, Ariel, Israel*

[b]*Ariel University, Chemical Engineering and Biotechnology Faculty, 40700, P.O.B. 3, Ariel, Israel*

[c]*Bar Ilan University, Physics Faculty, 52900, Ramat Gan, Israel*

[d]*Ariel University, Department of Electrical Engineering, 40700, P.O.B. 3, Ariel, Israel*

E-mail: edward@ariel.ac.il

**Abstract**

Plasma treatment increases the surface energy of condensed phases: solids and liquids. Two independent methods of the quantification of the influence imposed by a cold radiofrequency air plasma treatment on the surface properties of silicone oils (polydimethylsiloxane) of various molecular masses and castor oil are introduced. Under the first method the water droplet coated by oils was exposed to the cold air radiofrequency plasma, resulting in an increase of oil/air surface energy. An expression relating the oil/air surface energy to the apparent contact angle of the water droplet coated with oil was derived. The apparent contact angle was established experimentally. Calculation of the oil/air surface energy and spreading parameter was carried out for the various plasma-treated silicone and castor oils. The second method is based on the measurement of the electret response of the plasma-treated liquids.

**Keywords**: cold plasma; silicone oils; spreading parameter; change in the surface energy; hydrophilization; electret.

## 1. Introduction.

Plasma treatment (low and atmospheric-pressure) is widely used for the modification of surface properties of solid organic materials. [1] The plasma treatment creates a complex mixture of surface functionalities which influence surface physical and chemical properties; this results in a dramatic change in the wetting behaviour of the surface.[2] It was suggested that hydrophilization of organic surfaces by plasmas may be at least partially related to the re-orientation of hydrophilic moieties constituting organic molecules. [2i, 3-4] Oxidation of plasma-treated surfaces and removal of low-mass weight fragments present on organic surfaces also contribute to hydrophilization.[4] Much effort has been spent in understanding the interaction of plasmas with solid organic surfaces, whereas data related to plasma treatment of liquids are scarce.[5]

An interest in plasma treatment of liquid organic surfaces arose due to various practical reasons, including the possibility of de-contamination of liquids by plasmas[5] and microfluidics applications of these surfaces, stipulated by the extremely low contact angle hysteresis exhibited by liquid/liquid systems including oil/water ones.[6] Quantification of the impact exerted by plasmas on liquid surfaces faces serious experimental challenges.[7] Our paper focuses on the modification of surfaces of organic liquids exposed to cold radiofrequency plasma by two independent techniques, namely the measurement of the apparent contact angle and study of their electret response.

## 2. Experimental.

Quantification of the interaction of cold plasma with organic liquid surfaces was carried out with silicone-impregnated micro-porous polymer surfaces, manufactured as described in Refs. 8. Polypropylene (PP) films (the thickness 25 μm) were coated with honeycomb polycarbonate (PC) films, by the fast dip-coating process. As a result, we obtained typical "breath-figures" self-assembly patterns, depicted in **Figure 1**. Honeycomb PC coating was obtained according to the protocol described in detail in Refs. 8. The average radius of pores was about 1.5 μm. The average depth of pores as established by AFM was about 1 μm.

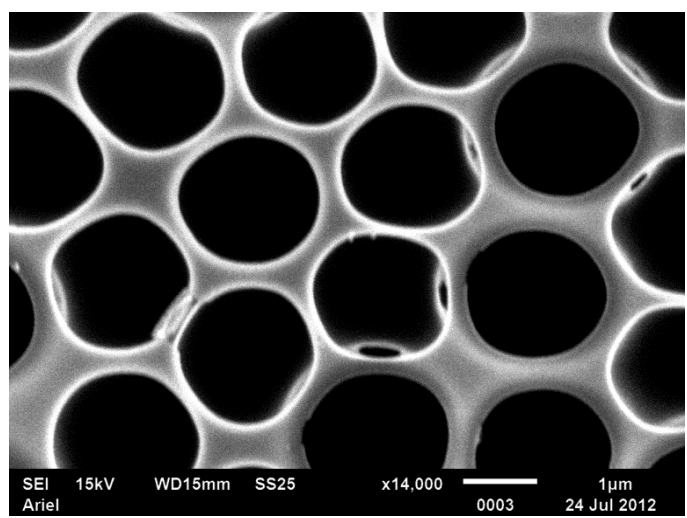

**Figure 1**. Polycarbonate honeycomb coating of polypropylene film , obtained with "breath-figures" self-assembly, carried out in a humid atmosphere. Scale bar is 1 μm.

Polydimethylsiloxane (PDMS) oils with molecular masses of 5600, 17500, 24000 g·mol$^{-1}$, and silicone oil for MP & BP apparatus (the molar mass was not identified), denoted for brevity in the text respectively as PDMS1, PDMS2, PDMS3 and PDMS4 were supplied by Aldrich. Castor oil was supplied by Vitamed Pharmaceutical Industries Ltd. PC porous coatings were impregnated by all kinds of aforementioned PDMS oils and castor oil, as shown in **Figure 2**. The thickness of PDMS/Castor oil layers was established by weighing as 20±2 μm.

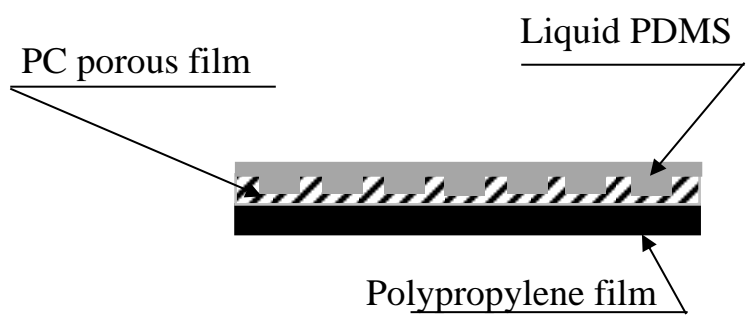

**Figure 2**. Scheme of the silicone oil impregnated PP substrates used in the investigation.

A water droplet with a volume of 8 μl was deposited on surfaces impregnated with the abovementioned oils as depicted in **Figure 2**. The water droplet was encapsulated by silicone and Castor oils, as depicted in **Figure 3** and as discussed in detail in Ref. 7.

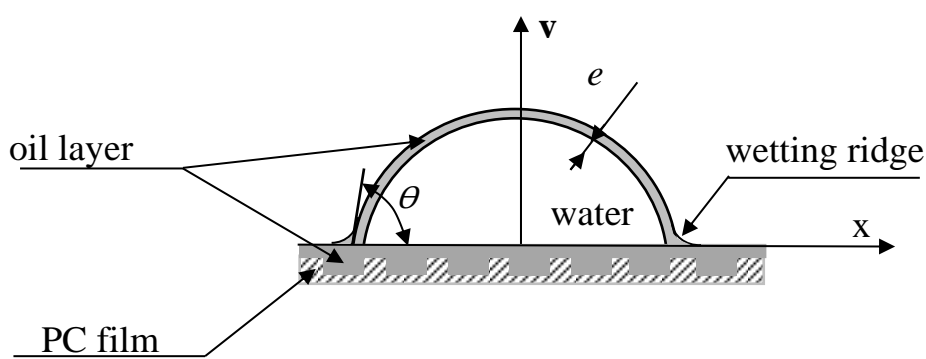

**Figure 3**. Water droplet encapsulated by the oil layer. $\theta$ is the apparent contact angle. $e$ is the thickness of the oil layer

Water droplets, encapsulated by all kinds of the oils used in our study, were exposed to a radiofrequency (13.56MHz) inductive air plasma discharge under the following parameters: pressure 266 Pa, power 18 W, ambient temperature. It should be stressed that the encapsulation of water droplets by oils prevented their evaporation in the plasma vacuum chamber, which made the entire experiment possible. The time of plasma treatment was varied within 30-60ms; the duration of a single pulse was 30ms.

Apparent contact angles of the water droplets encapsulated with the above oils were measured before and after the plasma treatment by a Ramé-Hart Advanced Goniometer (Model 500-F1).

Surface tensions of oils used in our investigation were established with the pendant droplet method by a Ramé-Hart Advanced Goniometer (Model 500-F1).

The electret response of oils exerted to the plasma treatment was studied with lab-made device depicted in **Figure 4**. PP films coated by PC porous films, filled by silicone and castor oils were exposed to cold plasma under the aforementioned parameters. Plasma treated samples were placed between two copper plates, mounted into a faradaic cage (see **Figure 4**).

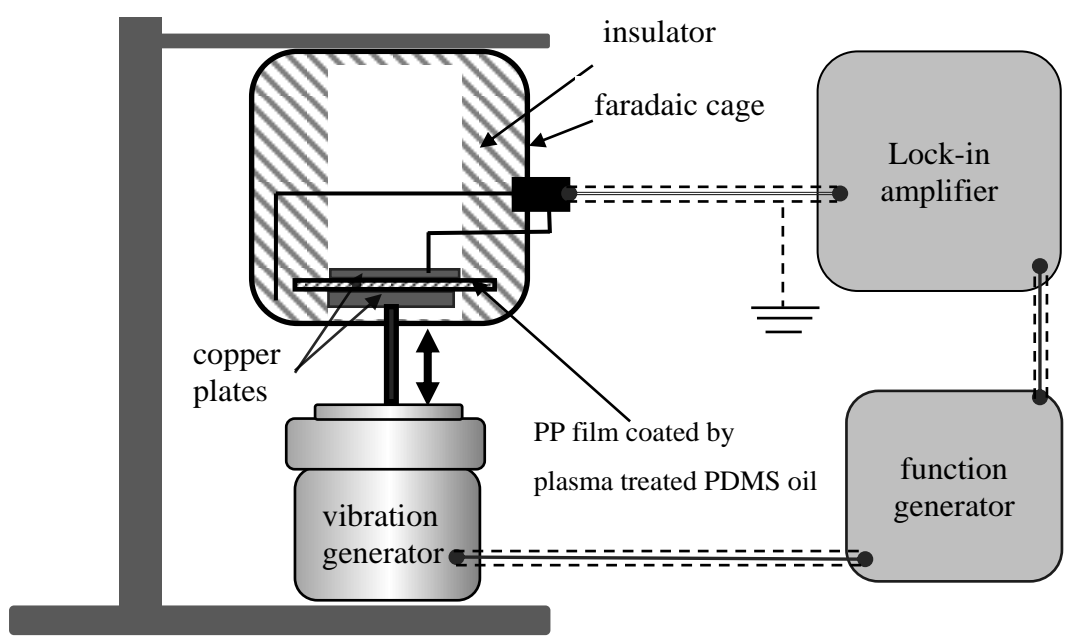

**Figure 4**. Scheme of the experimental lab-made unit for measuring of electret response of plasma-treated silicone oils.

The use of PP films possessing excellent dielectric properties in these experiments prevented possible artifacts due to the short circuiting across the sample. The lower plate was accelerated periodically by a vibration generator, therefore during the measurement the film was loaded with a dynamic oscillating force. The output signal of voltage between the charged surfaces was registered by a lock-in amplifier (7260 DSP). The vibration generator and lock-in amplifier were modulated by a function generator (Model DS345) at the frequency 107.77 Hz. Time dependencies of the electrical potential measured on the surface of the wet side of the samples were registered.

### 3. Results and discussion.

### 3.1 Influence of plasma treatment on the wetting regimes inherent for PDMS.

Water droplets deposited on silicone oils were encapsulated with them, as demonstrated in Refs. 7, 9. This wetting situation is well explained by the analysis of the spreading parameter $S$ governing the wetting situation:[10]

$$S = \gamma - (\gamma_{oil} + \gamma_{oil/water}),\qquad(1)$$

where $\gamma, \gamma_{oil}$ and $\gamma_{oil/water}$ are interfacial tensions at water/vapor, oil/vapor and oil/water interfaces respectively. Interfacial oil/water tensions for the studied oils, summarized in Table 1, were extracted from the literature data. [11]

Table 1. Interfacial properties of oils used in the investigation.

| Oil | Surface tension, $\gamma_{oil}$, mJ·m$^{-2}$ | Interfacial tension, $\gamma_{oil/water}$ mJ·m$^{-2}$ |
|---|---|---|
| PDMS1 | 20.7±0.3 | 23-24 |
| PDMS2 | 21.2±0.3 | 23-24 |
| PDMS3 | 21.1±0.5 | 23-24 |
| PDMS4 | 21.1±0.5 | 23-24 |
| Castor oil | 35.4±0.2 | 14.8 |

Substituting the aforementioned values of interfacial tensions in Equation (1), we obtain $S > 0$; in this case, both the silicone and castor oils are expected to completely coat the water droplet. This was indeed observed: water droplets were coated by silicone and castor oils.

We have already demonstrated, that cold plasma treatment increased the surface energy of silicone liquids. [7] The natural measure of the surface energy is an apparent contact angle, so-called APCA.[10] When the effects of disjoining pressure $P(e)$ and wetting ridge (see **Figure 3**) are neglected, the APCA for oils/water pairs in the situation of complete coating of a water droplet by oil, is given by the following expression:[7]

$$\cos\theta = \frac{\gamma_{oil} - \gamma_{oil/water}}{\gamma_{oil} + \gamma_{oil/water}}, \qquad (2)$$

The contact angle $\theta$ appearing in Equation (2) is an equilibrium contact angle.[7, 10] First of all we made sure that the APCAs observed for the studied oil/water pairs demonstrated properties inherent for equilibrium contact angles. Equilibrium contact angles do not depend on external fields, including gravity.[10, 12] Hence they are independent of the volume of a droplet. [10] We varied the volume of water droplet in the range of 3-8 µl with a step of 1 µl for all investigated water/oils pairs, and proved that the established APCAs are independent of the volume of the water droplets within the experimental accuracy of the measurement. Thus, it was reasonable to suggest that the established APCAs are the equilibrium ones.

Equation (2) may be exploited for the calculation of the surface energy of the oil. Rewriting of Equation (2) immediately yields:

$$\gamma_{oil} = \gamma_{oil/water} \cot^2 \frac{\theta}{2}, \qquad (3)$$

Thus, measurements of APCAs allow calculation of $\gamma_{oil}$ according to Equation (3), when $\gamma_{oil/water}$ is known from the independent measurements, such as those carried out in Refs. 11. If the value of $\gamma_{oil/water}$ is not affected by plasma treatment and $\theta$ is established experimentally, Equation (3) immediately supplies the value of $\gamma_{oil}$. The idea that $\gamma_{oil/water}$ was not essentially modified by plasma treatment seems reasonable, due to the fact that cold plasmas affect only the external nanometrically scaled layer of a surface.[1-2]

It should be emphasized that Equation (3) permits direct calculation of $\gamma_{oil}$ from the value of the APCA, because the APCA depends on the pair $\gamma_{oil}, \gamma_{oil/water}$ and not on the triad of interface tensions as it occurs for wetting of solid surfaces, described by the famous Young formula.[10,12]

As it is seen from the data, summarized in Table 2, the APCA decreased markedly with the time of plasma treatment for all kinds of oils used in our investigation. It should be stressed that in the present paper the time span of the plasma treatment was controlled precisely with a step of 30 ms. The decrease of APCA is also illustrated in **Figure 5**.

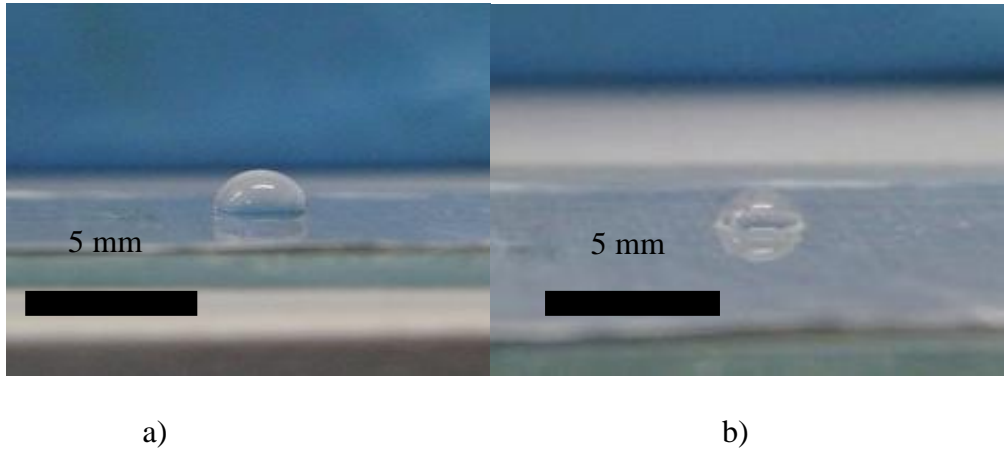

a)                                                    b)

**Figure 5**. Images of a water droplet coated with PDMS2 silicone oil: 0). before plasma treatment; b)
the same droplet after 60 ms of the cold plasma treatment. In both cases the water droplet is coated
by PDMS2 oil.

We observed the gradual decrease of the volume of the water droplet in the course of plasma
treatment. However this change in the droplet volume does not prevent the calculation of $\gamma_{oil}$
according to Equation (3), due to the fact that the equilibrium APCA is insensitive to the volume of a
droplet, as dicussed above.[10]

As it may be expected, $\gamma_{oil}$ increased with the time of plasma treatment, as it is clearly seen from
data supplied in Table 2.

Table 2. Dependence of APCA, $\theta$, surface tension $\gamma_{oil}$, and spreading parameter $S$ on the time of plasma treatment.

| Oil | Time of plasma treatment, ms | Contact angle $\theta$, degrees | $\gamma_{oil}$, mJ·m$^{-2}$ | $S$, mJ·m$^{-2}$ |
|---|---|---|---|---|
| PDMS 2 | 0 | 95 | 20.7 | 28.3 |
| | 30 | 73 | 42.8 | 6.1 |
| | 60 | 69 | 48.7 | 0.2 |
| PDMS 3 | 0 | 95 | 21.2 | 27.7 |
| | 30 | 87 | 25.5 | 23.4 |
| | 60 | 77 | 36.4 | 12.6 |
| PDMS 4 | 0 | 96 | 21.1 | 27.8 |
| | 30 | 73 | 42.0 | 6.9 |
| | 60 | 70 | 46.9 | 2.0 |
| Castor oil | 0 | 70 | 35.4 | 21.7 |
| | 30 | 61 | 42.7 | 14.5 |
| | 60 | 54 | 57.0 | 0.1 |

It was also instructive to deduce the dependence of the spreading parameter $S$ on the time of plasma treatment. The spreading parameter $S$ was calculated with Equation (1) for values of $\gamma_{oil}$, and estimated with Equation (3) from the experimental data, supplied in Table 2. It is seen, that the spreading parameter $S$ decreases with growth of the time of plasma treatment and approaches zero. This is quite understandable, when the increase in the oil/air surface energy arising from the plasma treatment is taken into account. When the spreading parameter approaches zero, the continuation of plasma treatment becomes impossible; silicone oil does not coat a water droplet and it is evaporated in the plasma chamber.

The maximal value of $\gamma_{oil}$ established for all kinds of silicone oils used in our study was close to $\gamma_{oil} = 49\,\text{mJ} \cdot \text{m}^{-2}$. It should be emphasized that this is not the "saturation value" of the oil/air interfacial tension. The saturation value of $\gamma_{oil}$ remains unachievable under the applied technique, due to the decrease of the spreading parameter finally approaching zero, resulting in the evapotation of a droplet, as explained above.

The important advantage of the proposed method of establishment of the surface energy of oils is stipulated by the low contact angle hysteresis, inherent for liquid/liquid systems, associated with the weak pinning of the contact line, intrinsic for these kinds of systems.[6]

### 3.2. Electret response of plasma-treated oils.

As it was demonstrated in the previous section plasma treatment increases the surface energy of organic oils. But what is the physical mechanism responsible for this increase? It is well accepted that the plasma treatment creates a complex mixture of surface functionalities which influence physical and chemical properties of a polymer surface, resulting in the jump in its surface energy.[2] In our paper we focus on one of the possible sources of this increase, namely the modification of electrical properties of oils by the plasma. Recent investigations demonstrated that plasmas modify essentially electrical properties of solid polymers including PDMS, PP and polyethylene.[13] We studied the electret response of the plasma-treated oils with the device, shown in **Figure 4**, already exploited by our group for the study of the electret response of plasma-treated solid polymers.[13b] The literature data related to the electret properties of liquids are scanty (an electret is a material that has a macroscopic electric field at its surface).[14] It should be emphasized that all studied organic oils demonstrated the pronounced electret response. The initial voltage $V_0$, established immediately after the plasma treatment, registered for plasma-treated oils was on the order of magnitude of 100 µV. However, this voltage decays with times, as shown in **Figures 6**a-b,

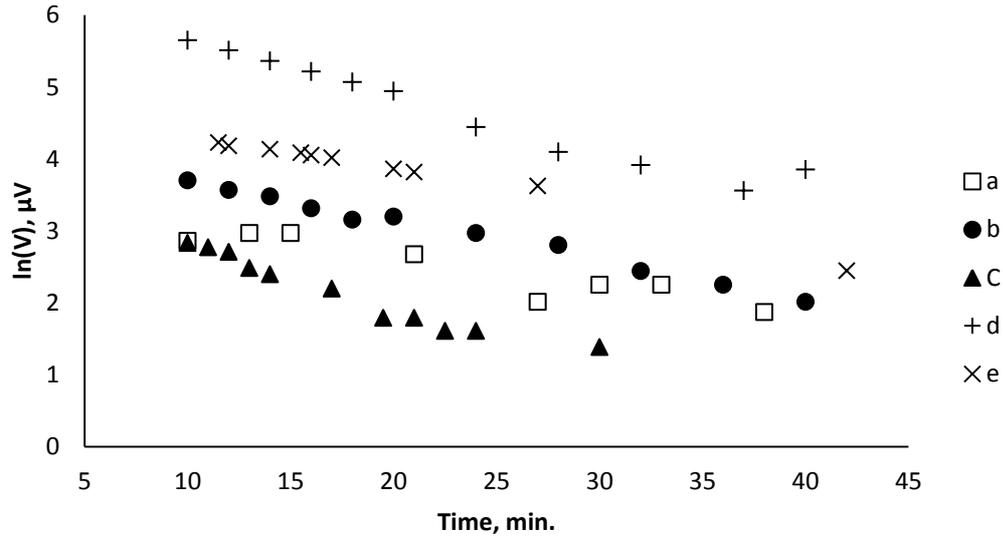

a.

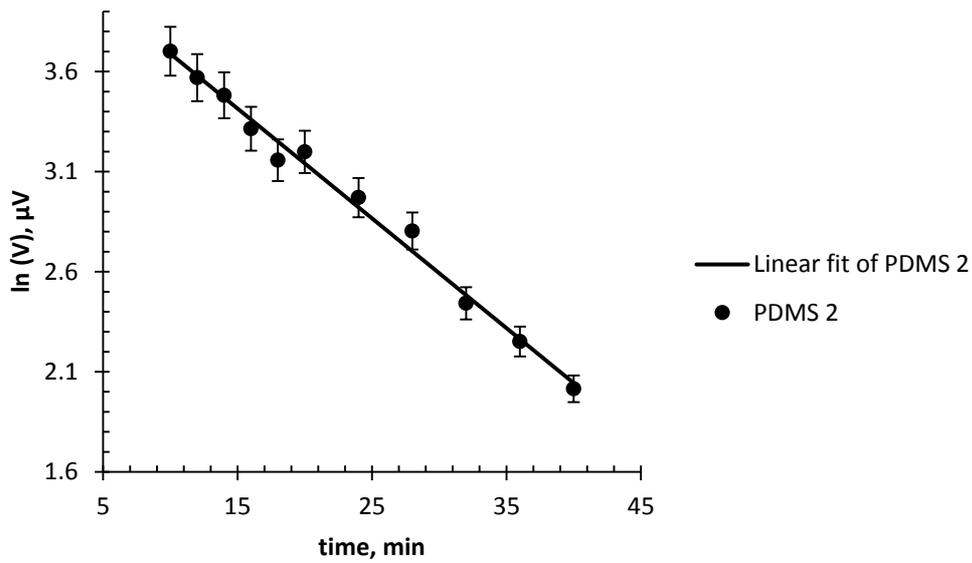

b.

**Figure 6**. a) Semi-*ln* dependence of the electret response (voltage) on time: a - PDMS1, b - PDMS 2, c - PDMS 3, d – PDMS 4, e – castor oil; b) The example of the linear fitting of the semi-*ln* dependence of the voltage on time for PDMS 2. Parameters of fitting are

$\ln(V) = (4.24 \pm 0.04) + \dfrac{t}{18.2 \pm 0.6[\min]}$; the squared coefficient of correlation for the fitting is R$^2$ = 0.989.

as it takes place on plasma-treated solid polymers.[13b] The temporal decay of voltage may be exponentially fitted by exponential dependence:

$$V(t) = V_0 + \tilde{V} \exp(-\frac{t}{\tau_e}), \qquad (4)$$

where $\tau_e$ is the characteristic time of the decay of electret properties, $\tilde{V}$ is the fitting parameter. The characteristic times established for investigated organic oils are summarized in Table 3.

Table 3. Characteristic times of the electret reponse $\tau_e$ decay established for various organic oils.

| Oil | $\tau_e$, ±2.5 min |
| --- | --- |
| PDMS1 | 18.0 |
| PDMS2 | 17.4 |
| PDMS3 | 12.5 |
| PDMS4 | 12.8 |
| Castor oil | 19.9 |

The open question to be addressed in future investigations: what are the microscopic sources of both of the hydrophilization of oils (discussed in the previous Section) and electret behavior of the oils observed on our experiments and illustrated with **Figure 6**a-b).

Generally, the observed behavior may be related to following effects: orientation of groups of polymer chains by plasma,[15] resulting in a non-zero dipole moment, oriented perpendicular to both the chain axis and the plane of the films, as it occurs with the poled PVDF,[16] and the charging of polymer films by the plasma. [13, 17] It is also possible that both of these effects act in parallel.

We incline to suggest that the charging of the surface by plasma is the main mechanism responsible for the modification of the electrical properties of a surface. Indeed, consider first the interaction of dipole groups of polymer chains with the electrical field of the plasma. Assume that the surface is built of moieties possessing the dipole moment $\vec{p}$. It seems reasonable to relate  relate at least partially the hydrophilization of the surface to the orientation of these moieties by the electric field of the plasma sheath $\vec{E}_s$, formed in the vinicity of a treated liquid.[18] The dimensionless parameter $\varsigma$, describing the interaction of the dipoles constituting the surface with the electric field of the plasma $\vec{E}_s$, is given by [19]:

$$\varsigma = \frac{pE_s}{k_B T},\qquad(5)$$

where $T$ is the temperature. The upper value of the achievable electric field of the plasma sheath may be estimated as $E_s^{\max} \cong \frac{\Phi_s}{\lambda_{De}} \cong \frac{100\,\text{V}}{10^{-4}\,\text{m}} = 10^6\,\text{V}\cdot\text{m}^{-1}$, where $\Phi_s \approx 100\,\text{V}$ is the potential of the liquid surface, and $\lambda_{De} \approx 10^{-4}\,\text{m}$ is the Debye length of the cold plasma.[18] Substituting $p \cong 1\text{D} \cong 3.3 \times 10^{-30}\,\text{C}\cdot\text{m}$, which is typical for moieties constituting polymer surfaces,[20] we obtain the upper estimation of $\varsigma$ for room temperatures: $\varsigma \cong 10^{-3}$. This means that the observed hydrophilization of liquid surfaces by cold plasmas could hardly be related to orientation of the dipole moieties forming the surface by the electrical field of a plasma sheath. This orientation will rapidly be destroyed at ambient conditions by the thermal agitation of dipole groups. Hence, the reasonable mechanism of the modification of liquid surfaces by plasma should be related to the charging of the surface by plasma.[13] The kinetic model of charging was proposed recently in Ref. 21.

Electrical charge gained by a liquid surface is lost with time. The similarity of the electret response and hydrophobic recovery time scales[7] hints that the physical processes responsible for hydrophobic recovery and decay of the electret response are generally the same. However, the precise identification of these processes remains quite challenging.

## 4. Conclusions

We conclude that measurements of the apparent contact angle and the electret response could be exploited for the establishment of the surface energy of liquids, in the situation where the spreading parameter is positive. We tested these methods with water/silicone and water/castor oil systems. Silicone and castor oils coated and encapsulated the water droplets. This enabled exposure of water/silicone oil "sandwiches" to the cold radiofrequency air plasma. We demonstrated recently that radiofrequency plasma increases the surface energy of the liquid polymers. An expression relating the oil/air surface energy to the apparent contact angle of a water droplet coated with oil was derived. This expression allowed quantification of the impact exerted by cold plasma treatment on the surface energy of the oil/air interface. Calculation of the oil/air surface energy and spreading parameter was carried out for different time spans of plasma treatment for silicone oils possessing various molecular weights and also for the castor oil.

The increase in the surface energy of plasma-treated organic liquid surfaces may be at least partially related to the charging of liquid surface by plasma. Plasma-treated liquid surfaces demonstrated a strong electret response, decaying with time. The similarity of the electret response and hydrophobic recovery time scales leads to the conclusion that physico-chemical processes responsible for a hydrophobic recovery and decay of the electret response are generally the same.

### Acknowledgements

We are thankful to Mrs. Y. Bormashenko for her help in preparing this manuscript. The Authors are thankful to Mrs. N. Litvak for SEM imaging of PC substrates. The work was partially supported by the ACS Petroleum Research Fund (Grant 52043-UR5).